\documentclass[10pt,conference]{IEEEtran}
\IEEEoverridecommandlockouts
\usepackage{cite}
\usepackage{amsmath,amssymb,amsfonts}
\usepackage{algorithmic}
\usepackage{graphicx}
\usepackage{textcomp}
\usepackage{multirow}
\usepackage{booktabs}
\usepackage{float}
\usepackage[table,xcdraw]{xcolor}
\def\BibTeX{{\rm B\kern-.05em{\sc i\kern-.025em b}\kern-.08em
    T\kern-.1667em\lower.7ex\hbox{E}\kern-.125emX}}
\begin{document}

\title{RIS-Aided Mobile Network Design\\
\thanks{This work was completed as part of project no. $2021/43/\text{B}/\text{ST}7/01365$, funded by the National Science Centre (NSC) in Poland.}
}

\author{\IEEEauthorblockN{Adam Samorzewski}
\IEEEauthorblockA{\textit{Institute of Radiocommunications} \\
\textit{Poznan University of Technology}\\
Poznan, Poland \\
adam.samorzewski@put.poznan.pl}
\and
\IEEEauthorblockN{Adrian Kliks}
\IEEEauthorblockA{\textit{Institute of Radiocommunications} \\
\textit{Poznan University of Technology}\\
Poznan, Poland \\
adrian.kliks@put.poznan.pl}
}

\maketitle

\begin{abstract}
In this paper, we examine the distribution of radio signal propagation within the city of Poznan (Poland) to determine optimal locations for deploying Reconfigurable Intelligent Surfaces (RIS). The study focuses on designing a 5G/6G Radio Access Network (RAN), incorporating eight Base Stations (BSs) that utilize either Single Input Single Output (SISO), or Multiple Input Multiple Output (MIMO) antenna technologies, depending on the network cell configuration. Through detailed simulations and analyses, we explore various propagation scenarios in both Line-of-Sight (LOS) and Non-Line-of-Sight (NLOS) conditions, considering the complex urban landscape characterized by high-rise buildings. The results demonstrate the potential of using RISs in mobile networks to enhance radio signal quality in urban environments through strategic placements. Our findings suggest that RISs can significantly mitigate Path Loss (PL) and improve signal coverage in challenging urban environments, particularly in areas where traditional base station deployment alone would be insufficient. Furthermore, the study highlights the role of RISs in reducing the need for additional base stations, thereby optimizing network costs and infrastructure while maintaining high-quality service delivery. The insights gained from this research provide valuable guidelines for network planners and engineers seeking to implement RIS technology in future 5G and beyond networks, ensuring more efficient and robust urban communication systems.\footnote{Copyright © 2025 IEEE. Personal use is permitted. For any other purposes, permission must be obtained from the IEEE by emailing pubs-permissions@ieee.org. This is the author’s version of an article that has been published in the proceedings of the 2025 IEEE 36th International Symposium on Personal, Indoor and Mobile Radio Communications (PIMRC) by the IEEE. Changes were made to this version by the publisher before publication, the final version of the record is available at: https://dx.doi.org/10.1109/PIMRC62392.2025.11274762. To cite the paper use: A. Samorzewski, A.~Kliks, “Signal Propagation in RIS-Aided 5G Systems,” in: \textit{2025 IEEE 36th International Symposium on Personal, Indoor and Mobile Radio Communications (PIMRC)}, Istanbul, Türkiye, 2025, pp.~1--6, doi: 10.1109/PIMRC62392.2025.11274762 or visit https://ieeexplore.ieee.org/document/11274762.}
\end{abstract}


\section{Introduction}
\label{section:introduction}
In recent years, Reconfigurable Intelligent Surfaces have emerged as a revolutionary technology with the potential to reshape wireless communication systems. Unlike conventional antenna technologies that depend on beamforming and power control at the transmitter, RISs introduce a new approach by directly modifying the propagation environment. These surfaces consist of large arrays of passive (or active) elements, such as metamaterials or simple reflectors, that can dynamically change the electromagnetic properties of the surroundings to improve signal transmission and reception. RIS technology builds upon principles of smart and reflective surfaces, incorporating advanced signal processing and optimization techniques to control electromagnetic waves in real time. By carefully adjusting the phase shifts and amplitude of reflected signals, RISs can focus and direct beams, counteract channel impairments, and reduce path losses in various propagation conditions. This ability allows Mobile Network Operators (MNOs) to manage radio signals effectively, ensuring coverage in areas where reception is challenging. Additionally, RIS arrays can significantly reduce inter-signal interference and minimize users' exposure to electromagnetic fields (EMF). These arrays are especially beneficial in urban environments, where transmitted signals are frequently refracted and scattered by numerous obstacles, such as buildings. In rural regions, active RIS devices that amplify redirected signals could serve as relays, decreasing the number of base stations required to provide coverage in the area \cite{Huang, DiRenzo, Tang, Liu, SamorzewskiJTIT2023, SamorzewskiKRiT2023, SamorzewskiSoftCOM2023, SamorzewskiGLOBECOM2023}.


This paper analyzes the use of Reconfigurable Intelligent Surface (RIS) matrices in an urban environment, focusing on their impact on the path loss (PL) characteristics of propagated radio signals. Additionally, the study examines how the suspension of RIS influences coverage within the studied area.

The article is organized as follows: Section \ref{section:scenario} outlines the system scenario under consideration; Section \ref{section:simulation} details the research methodology and simulation configurations; Section \ref{section:results} presents the results through graphs and tables, highlighting the average values of the observed system parameters; finally, Section \ref{section:conclusions} concludes the paper by summarizing the key findings derived from the results.

\section{Considered Scenario}
\label{section:scenario}
\begin{figure}[htb]
\centering
\includegraphics[width=0.49\textwidth]{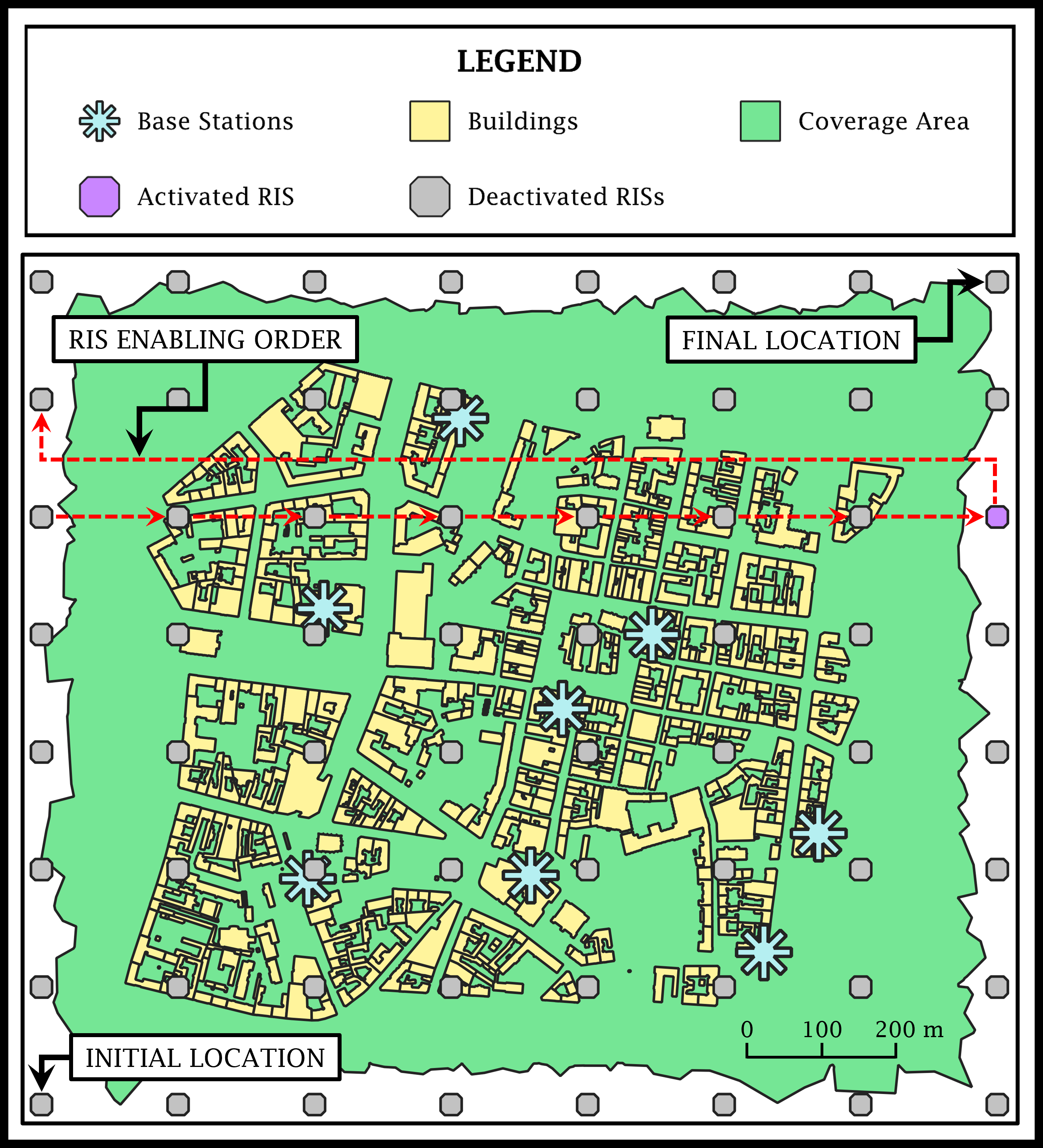}
\caption{Research methodology schema}
\label{figure:methodology_schema}
\end{figure}

The scenario configured for this study, as illustrated in Fig.~\ref{figure:methodology_schema}, focuses on an urban area, specifically the neighborhood of the Old Market in Poznan. The data used to define the coverage area and building layouts are real datasets provided by the Poznan City Hall \cite{AreaData}.

The assumed mobile network comprises eight base stations, each with three network cells operating at $800$ MHz, $2100$ MHz, and $3500$ MHz. The BS locations were selected based on real data provided by one of the major Polish mobile operators \cite{NetworkData}. The antenna systems within these cells utilize either SISO technology at $800$ and $2100$ MHz or MIMO technology at $3500$ MHz. The SISO systems employ a single active element per transceiver, while the MIMO systems use $64$ active elements per transceiver \cite{Castellanos, SamorzewskiGLOBECOM2023}.

Additionally, $64$ potential locations for deploying RIS were identified within the study area. The placement selection sequence starts from the bottom-left corner of the grid, proceeding left to right, and then moving up to the next row, continuing this pattern until reaching the top-right corner. These locations were automatically selected by the simulation software after specifying the number of rows and columns for the RIS grid.

For each RIS location, path loss distribution across the area is measured. Four result maps are generated: the minimal PL distribution with RIS disabled and enabled, the averaged PL distribution with RIS activated, and the distribution for RIS-BS signal reinforcement. Based on these maps, additional results, including Cumulative Distribution Functions (CDFs) and summary tables, are prepared to further analyze the system's performance.

\section{Simulation Configuration}
\label{section:simulation}
The research was conducted through simulations using specialized software known as the Green Radio Access Network Design (GRAND) tool \cite{Castellanos, SamorzewskiGLOBECOM2023}.

The primary objective of the study was to analyze the distribution of radio signal path loss in the selected urban area, both with and without the deployment of RIS placed in different locations (to find the most beneficial ones). To evaluate radio signal attenuation across the area, the following propagation models were utilized:
\begin{itemize}
\item $3$GPP TR $38.901$ UMa (Urban Macro) \textbf{--} signal propagation in LOS and NLOS scenarios, i.e., radio waves distributed {\it directly} from a base station \cite{3GPP},
\item RIS-FFBC (RIS Far Field Beamforming Case) \textbf{--} signal propagation by a base station via reflecting radio waves from a Reconfigurable Intelligent Surface~\cite{Tang}.
\end{itemize}

The simulation setup comprises $24$ repetitions, with each run dedicated to estimating the path loss distribution within the study area, both with and without the RIS positioned at a preselected location. The parameters configured for the simulation are detailed in Tab.~\ref{table:configuration_network_design} and \ref{table:configuration_ris_design}.

\begin{table}[t]
\centering
\caption{Parameters configuration (BS)
\cite{Castellanos, SamorzewskiGLOBECOM2023, 3GPP, Bjornson, AreaData, NetworkData}
}
\label{table:configuration_network_design}
    \resizebox{0.49\textwidth}{!}{\begin{tabular}{|l|c|c|c|cccc|}
    \arrayrulecolor[HTML]{002060}\hline
    \multirow{3}{*}{\cellcolor[HTML]{002060}}   & \multirow{3}{*}{\cellcolor[HTML]{002060}{\color[HTML]{FFFFFF}}} & \multirow{3}{*}{\cellcolor[HTML]{002060}{\color[HTML]{FFFFFF} }} & \multirow{3}{*}{\cellcolor[HTML]{002060}{\color[HTML]{FFFFFF} }} & \multicolumn{3}{c|}{\cellcolor[HTML]{002060}{\color[HTML]{FFFFFF} Value}}                                                                                                                                        \\ \cline{5-7} 
                        \cellcolor[HTML]{002060}{\color[HTML]{FFFFFF}} & \cellcolor[HTML]{002060}{\color[HTML]{FFFFFF} Parameter}                           &  \cellcolor[HTML]{002060}{\color[HTML]{FFFFFF} Sign}                     &  \cellcolor[HTML]{002060}{\color[HTML]{FFFFFF} Unit}                      & \multicolumn{3}{c|}{\cellcolor[HTML]{FFFFFF}{\color[HTML]{000000} \textit{Network Cell}}} \\ \cline{5-7}
                        \cellcolor[HTML]{002060}{\color[HTML]{FFFFFF}} & \cellcolor[HTML]{002060}{\color[HTML]{FFFFFF}}                           & \cellcolor[HTML]{002060}{\color[HTML]{FFFFFF}}                      & \cellcolor[HTML]{002060}{\color[HTML]{FFFFFF}}                      & \multicolumn{1}{c|}{$1\text{.}$}    & \multicolumn{1}{c|}{$2\text{.}$}   & \multicolumn{1}{c|}{$3\text{.}$} \\ \hline
    \multirow{1}{*}{\cellcolor[HTML]{89B0FF}{\color[HTML]{002060}}} 
                        \cellcolor[HTML]{89B0FF}{\color[HTML]{002060}} & \cellcolor[HTML]{E7E7E7}{\color[HTML]{000000} Quantity}                   & \cellcolor[HTML]{E7E7E7}{\color[HTML]{000000}$K_\text{BS}$}            & \cellcolor[HTML]{E7E7E7}{\color[HTML]{000000}\textbf{--}}                     &                                    \multicolumn{3}{c|}{\cellcolor[HTML]{E7E7E7}{\color[HTML]{000000}$8$}}      \\ \cline{2-7} 
                        \multirow{-2}{*}{\rotatebox[origin=c]{90}{\cellcolor[HTML]{89B0FF}{\color[HTML]{002060} Gen.}}} & Technology & \textbf{--} & \textbf{--} & \multicolumn{3}{c|}{$5\text{G}$} \\ 
                        \hline
    \multirow{1}{*}{\cellcolor[HTML]{89B0FF}{\color[HTML]{002060}}}
                        \cellcolor[HTML]{89B0FF}{\color[HTML]{002060}} & \cellcolor[HTML]{E7E7E7}{\color[HTML]{000000} Frequency}                  & \cellcolor[HTML]{E7E7E7}{\color[HTML]{000000} $f$}            & \cellcolor[HTML]{E7E7E7}{\color[HTML]{000000}{$\left[\text{MHz}\right]$}}             & \multicolumn{1}{c|}{\cellcolor[HTML]{E7E7E7}{\color[HTML]{000000} $\quad800\quad$}}  & \multicolumn{1}{c|}{\cellcolor[HTML]{E7E7E7}{\color[HTML]{000000}$\quad2100\quad$}} & \multicolumn{1}{c|}{\cellcolor[HTML]{E7E7E7}{\color[HTML]{000000}$\quad3500\quad$}} \\ \cline{2-7} 
                        \cellcolor[HTML]{89B0FF}{\color[HTML]{002060}} & Channel Bandwidth          & $B_\text{w}$            & {$\left[\text{MHz}\right]$}      & \multicolumn{1}{c|}{$80$}   & \multicolumn{1}{c|}{$120$}  & \multicolumn{1}{c|}{$120$} \\ \cline{2-7} 
                        \cellcolor[HTML]{89B0FF}{\color[HTML]{002060}} & \cellcolor[HTML]{E7E7E7}{\color[HTML]{000000} Used Subcarriers}           & \cellcolor[HTML]{E7E7E7}{\color[HTML]{000000} $N_\text{SC,u}$}            & \cellcolor[HTML]{E7E7E7}{\color[HTML]{000000} \textbf{--}}                     & \multicolumn{3}{c|}{\cellcolor[HTML]{E7E7E7}{\color[HTML]{000000} $320$}} \\ \cline{2-7} 
                        \cellcolor[HTML]{89B0FF}{\color[HTML]{002060}} & Total Subcarriers          & $N_\text{SC,t}$            & \textbf{--}                     & \multicolumn{3}{c|}{$512$} \\ \cline{2-7} 
                        \cellcolor[HTML]{89B0FF}{\color[HTML]{002060}} & \cellcolor[HTML]{E7E7E7}{\color[HTML]{000000} Sampling Factor}            & \cellcolor[HTML]{E7E7E7}{\color[HTML]{000000} SF}                    & \cellcolor[HTML]{E7E7E7}{\color[HTML]{000000} \textbf{--}}                     & \multicolumn{3}{c|}{\cellcolor[HTML]{E7E7E7}{\color[HTML]{000000} $1.536$}} \\ \cline{2-7} 
                        \cellcolor[HTML]{89B0FF}{\color[HTML]{002060}} & Pilot Reuse Factor         & RF                    & \textbf{--}                     & \multicolumn{3}{c|}{$1$} \\ \cline{2-7} 
                        \cellcolor[HTML]{89B0FF}{\color[HTML]{002060}} & \cellcolor[HTML]{E7E7E7}{\color[HTML]{000000} Coherence Time}             & \cellcolor[HTML]{E7E7E7}{\color[HTML]{000000} $t_\text{c}$}            & \cellcolor[HTML]{E7E7E7}{\color[HTML]{000000}{$\left[\text{ms}\right]$}}              & \multicolumn{3}{c|}{\cellcolor[HTML]{E7E7E7}{\color[HTML]{000000}$50$}} \\ \cline{2-7} 
                        \cellcolor[HTML]{89B0FF}{\color[HTML]{002060}} & Coherence Bandwidth        & $B_\text{c}$            & {$\left[\text{MHz}\right]$}             & \multicolumn{3}{c|}{$1$} \\ \cline{2-7}
                        \multirow{-9}{*}{\rotatebox[origin=c]{90}{\cellcolor[HTML]{89B0FF}{\color[HTML]{002060} Band}}} & \cellcolor[HTML]{E7E7E7}{\color[HTML]{000000} Spatial Duty Cycle}         & \cellcolor[HTML]{E7E7E7}{\color[HTML]{000000} $S$}            & \cellcolor[HTML]{E7E7E7}{\color[HTML]{000000}{$\left[\text{\%}\right]$}}              & \multicolumn{1}{c|}{\cellcolor[HTML]{E7E7E7}{\color[HTML]{000000} $0$}}    & \multicolumn{1}{c|}{\cellcolor[HTML]{E7E7E7}{\color[HTML]{000000} $0$}}    & \multicolumn{1}{c|}{\cellcolor[HTML]{E7E7E7}{\color[HTML]{000000} $25$}} \\ \hline
    \multirow{1}{*}{\cellcolor[HTML]{89B0FF}{\color[HTML]{002060}}}  & Antenna Height             & $h_\text{BS}$            & {$\left[\text{m}\right]$}               & \multicolumn{3}{c|}{$\left(27,47\right)$} \\ \cline{2-7} 
                        \cellcolor[HTML]{89B0FF}{\color[HTML]{002060}} & \cellcolor[HTML]{E7E7E7}{\color[HTML]{000000} Antenna Elements}           & \cellcolor[HTML]{E7E7E7}{\color[HTML]{000000} $M_\text{BS}$}            & \cellcolor[HTML]{E7E7E7}{\color[HTML]{000000} \textbf{--}}                     & \multicolumn{1}{c|}{\cellcolor[HTML]{E7E7E7}{\color[HTML]{000000} $1$}}    & \multicolumn{1}{c|}{\cellcolor[HTML]{E7E7E7}{\color[HTML]{000000} $1$}}    & \multicolumn{1}{c|}{\cellcolor[HTML]{E7E7E7}{\color[HTML]{000000} $64$}} \\ \cline{2-7} 
                        \cellcolor[HTML]{89B0FF}{\color[HTML]{002060}} & Antenna Gain               & $G_\text{a}$            & {$\left[\text{dBi}\right]$}             & \multicolumn{1}{c|}{$16$}   & \multicolumn{1}{c|}{$18$}   & \multicolumn{1}{c|}{$24$} \\ \cline{2-7} 
                        \cellcolor[HTML]{89B0FF}{\color[HTML]{002060}} & \cellcolor[HTML]{E7E7E7}{\color[HTML]{000000} Feeder Loss}        & \cellcolor[HTML]{E7E7E7}{\color[HTML]{000000} $L_\text{f}$}            & \cellcolor[HTML]{E7E7E7}{\color[HTML]{000000}{$\left[\text{dBi}\right]$}}             & \multicolumn{1}{c|}{\cellcolor[HTML]{E7E7E7}{\color[HTML]{000000} $2$}}    & \multicolumn{1}{c|}{\cellcolor[HTML]{E7E7E7}{\color[HTML]{000000} $2$}}    & \multicolumn{1}{c|}{\cellcolor[HTML]{E7E7E7}{\color[HTML]{000000} $3$}} \\ \cline{2-7} 
                        \cellcolor[HTML]{89B0FF}{\color[HTML]{002060}} & Transmit Power        & $P_\text{TX}$            & {$\left[\text{dBm}\right]$}             & \multicolumn{3}{c|}{$43$} \\ \cline{2-7} 
                        \multirow{-6}{*}{\rotatebox[origin=c]{90}{\cellcolor[HTML]{89B0FF}{\color[HTML]{002060} Transceivers}}} & \cellcolor[HTML]{E7E7E7}{\color[HTML]{000000} Noise Factor}               & \cellcolor[HTML]{E7E7E7}{\color[HTML]{000000} NF}                    & \cellcolor[HTML]{E7E7E7}{\color[HTML]{000000}{$\left[\text{dB}\right]$}}              & \multicolumn{1}{c|}{\cellcolor[HTML]{E7E7E7}{\color[HTML]{000000} $8$}}    & \multicolumn{1}{c|}{\cellcolor[HTML]{E7E7E7}{\color[HTML]{000000} $8$}}    & \multicolumn{1}{c|}{\cellcolor[HTML]{E7E7E7}{\color[HTML]{000000} $7$}} \\ \hline
    \multirow{1}{*}{\cellcolor[HTML]{89B0FF}{\color[HTML]{002060}}}  & Path Loss Model            & \textbf{--}                     & \textbf{--}                     & \multicolumn{3}{c|}{$3$GPP TR $38.901$ UMa \cite{3GPP}} \\ \cline{2-7} 
                        \cellcolor[HTML]{89B0FF}{\color[HTML]{002060}} & \cellcolor[HTML]{E7E7E7}{\color[HTML]{000000} Interference Margin}        & \cellcolor[HTML]{E7E7E7}{\color[HTML]{000000} IM}                    & \cellcolor[HTML]{E7E7E7}{\color[HTML]{000000}{$\left[\text{dB}\right]$}}              & \multicolumn{3}{c|}{\cellcolor[HTML]{E7E7E7}{\color[HTML]{000000} $2$}}  \\ \cline{2-7} 
                        \cellcolor[HTML]{89B0FF}{\color[HTML]{002060}} & Doppler Margin             & DM                    & {$\left[\text{dB}\right]$}              & \multicolumn{3}{c|}{$3$} \\ \cline{2-7} 
                        \cellcolor[HTML]{89B0FF}{\color[HTML]{002060}} & \cellcolor[HTML]{E7E7E7}{\color[HTML]{000000} Fade Margin}                & \cellcolor[HTML]{E7E7E7}{\color[HTML]{000000} FM}                    & \cellcolor[HTML]{E7E7E7}{\color[HTML]{000000}{$\left[\text{dB}\right]$}}              & \multicolumn{3}{c|}{\cellcolor[HTML]{E7E7E7}{\color[HTML]{000000} $10$}} \\ \cline{2-7} 
                        \cellcolor[HTML]{89B0FF}{\color[HTML]{002060}} & Shadow Margin              & SM                    & {$\left[\text{dB}\right]$}              & \multicolumn{1}{c|}{$12.8$} & \multicolumn{1}{c|}{$15.2$} & \multicolumn{1}{c|}{$10$} \\ \cline{2-7} 
                       \multirow{-6}{*}{\rotatebox[origin=c]{90}{\cellcolor[HTML]{89B0FF}{\color[HTML]{002060} Propagation}}} & \cellcolor[HTML]{E7E7E7}{\color[HTML]{000000} Implementation Loss}         & \cellcolor[HTML]{E7E7E7}{\color[HTML]{000000} IL}            & \cellcolor[HTML]{E7E7E7}{\color[HTML]{000000}{$\left[\text{dB}\right]$}}              & \multicolumn{1}{c|}{\cellcolor[HTML]{E7E7E7}{\color[HTML]{000000} $0$}}    & \multicolumn{1}{c|}{\cellcolor[HTML]{E7E7E7}{\color[HTML]{000000} $0$}}    & \multicolumn{1}{c|}{\cellcolor[HTML]{E7E7E7}{\color[HTML]{000000} $3$}} \\ \hline
    \end{tabular}}
\end{table}
\begin{table}[t]
\centering
\caption{Parameters configuration (RIS)
\cite{Tang}
}
\label{table:configuration_ris_design}
    \resizebox{0.39\textwidth}{!}{\begin{tabular}{|l|c|c|c|c|}
    \arrayrulecolor[HTML]{002060}\hline
                        \cellcolor[HTML]{002060}{\color[HTML]{FFFFFF}} & \cellcolor[HTML]{002060}{\color[HTML]{FFFFFF} Parameter}                           &  \cellcolor[HTML]{002060}{\color[HTML]{FFFFFF} Sign}                     &  \cellcolor[HTML]{002060}{\color[HTML]{FFFFFF} Unit}                      & \cellcolor[HTML]{002060}{\color[HTML]{FFFFFF} Value} \\ \hline
                        \multirow{1}{*}{\cellcolor[HTML]{89B0FF}{\color[HTML]{002060}}} 
                        \cellcolor[HTML]{89B0FF}{\color[HTML]{002060}} & \cellcolor[HTML]{E7E7E7}{\color[HTML]{000000} Locations' Quantity}                   & \cellcolor[HTML]{E7E7E7}{\color[HTML]{000000} $K_\text{RIS}$}            & \cellcolor[HTML]{E7E7E7}{\color[HTML]{000000}\textbf{--}}                     &                                    \multicolumn{1}{c|}{\cellcolor[HTML]{E7E7E7}{\color[HTML]{000000}$64$}}      \\ \cline{2-5} 
                        \multirow{-6}{*}{\rotatebox[origin=c]{90}{\cellcolor[HTML]{89B0FF}{\color[HTML]{002060}}}} & {\color[HTML]{000000} Column Elements}         & {\color[HTML]{000000} $M_\text{RIS}$}            & {\color[HTML]{000000}{\textbf{--}}}              & \multicolumn{1}{c|}{{\color[HTML]{000000} $102$}} \\ \cline{2-5} 
                        \cellcolor[HTML]{89B0FF}{\color[HTML]{002060}} & \cellcolor[HTML]{E7E7E7}{\color[HTML]{000000} Row Elements}        & \cellcolor[HTML]{E7E7E7}{\color[HTML]{000000} $N_\text{RIS}$}                    & \cellcolor[HTML]{E7E7E7}{\color[HTML]{000000}{\textbf{--}}}              & \multicolumn{1}{c|}{\cellcolor[HTML]{E7E7E7}{\color[HTML]{000000} $100$}}  \\ \cline{2-5} 
                        \cellcolor[HTML]{89B0FF}{\color[HTML]{002060}} & Column Width             & $d_\text{m}$                    & {$\left[\text{m}\right]$}              & \multicolumn{1}{c|}{$0.01$} \\ \cline{2-5} 
                        \cellcolor[HTML]{89B0FF}{\color[HTML]{002060}} & \cellcolor[HTML]{E7E7E7}{\color[HTML]{000000} Row Width}                & \cellcolor[HTML]{E7E7E7}{\color[HTML]{000000} $d_\text{n}$}                    & \cellcolor[HTML]{E7E7E7}{\color[HTML]{000000}{$\left[\text{m}\right]$}}              & \multicolumn{1}{c|}{\cellcolor[HTML]{E7E7E7}{\color[HTML]{000000} $0.01$}} \\ \cline{2-5} 
                        \cellcolor[HTML]{89B0FF}{\color[HTML]{002060}} & Reflected Signal Amp. Factor         & $A$                    & {\textbf{--}}              & \multicolumn{1}{c|}{$0.9$} \\ \cline{2-5}
                        \cellcolor[HTML]{89B0FF}{\color[HTML]{002060}} & \cellcolor[HTML]{E7E7E7}{\color[HTML]{000000} Suspension Height} & \cellcolor[HTML]{E7E7E7}{\color[HTML]{000000} $h_\text{RIS}$}                    & \cellcolor[HTML]{E7E7E7}{\color[HTML]{000000} $\left[\text{m}\right]$}              & \multicolumn{1}{c|}{\cellcolor[HTML]{E7E7E7}{\color[HTML]{000000} $40$}} \\ \cline{2-5}  
                        \multirow{1}{*}{\cellcolor[HTML]{89B0FF}{\color[HTML]{002060}}}  & {\color[HTML]{000000} Path Loss Model} & {\color[HTML]{000000} \textbf{--}}                     & {\color[HTML]{000000} \textbf{--}}     & \multicolumn{1}{c|}{{\color[HTML]{000000} RIS-FFBC \cite{Tang}}}\\ \hline
    \end{tabular}}
\end{table}

\section{Results}
\label{section:results}
In the following section, the results obtained for the test scenario are presented. First, the parameters that highlight the positive impact of RIS deployment on radio signal path loss distribution are reviewed. The results are then detailed in the form of tables (Tab.~\ref{table:results_pl_values} and \ref{table:results_gain}) and figures (Fig.~\ref{figure:pl_distribution} and \ref{figure:cdf_pl}), followed by a discussion of the findings.

\subsection{Methodology}
\label{subsection:methodology}
The study was conducted through simulation runs using dedicated software known as the GRAND tool \cite{Castellanos, SamorzewskiGLOBECOM2023}, developed in Java. This software is tailored for designing mobile networks based on input files that provide detailed information about the coverage area, the buildings within it, and the base stations, including the technologies employed. Additionally, the GRAND tool can define and integrate supporting equipment, such as Reconfigurable Intelligent Surfaces, by specifying the necessary input files. The simulation parameters, including the area layout, building distribution, and base station locations, were configured using real data provided by the Poznan City Hall \cite{AreaData} and a leading Polish mobile operator \cite{NetworkData}.

The GRAND tool is capable of evaluating radio signal path loss using mathematical models specified by the 3GPP (UMa \cite{3GPP}) and established in the scientific literature (RIS-FFBC \cite{Tang}). Using these models, the path loss distribution for the network environment under study was calculated for four different scenarios: with RIS disabled (labeled as BS), with RIS enabled and minimal path loss values selected at specific measurement locations (labeled as RIS), with RIS enabled where radio signals propagated via LOS or NLOS by base stations are amplified by signals reflected from RIS (labeled as RIS,BS), and with RIS enabled where the minimal path loss values (both \textit{directly} from BS and via RIS reflections) are averaged (labeled as AVG).

For this study, it was assumed that in each location, the RIS could rotate to maximize its impact on the path loss distribution. As a result, the elevation and azimuth angles between the RIS and a specific base station, as well as between the RIS and a particular Measurement Point (MP), were fixed at values of $\theta_\text{t}=\frac{\pi}{4}$, $\varphi_\text{t}=\pi$, $\theta_\text{r}=\frac{\pi}{4}$, and $\varphi_\text{r}=0$. These values were chosen based on the work described in \cite{Tang}, which demonstrated that maximum received power from the radio signal reflected by RIS could be achieved by adopting these specific angle values. Additionally, due to the use of the RIS path loss model (RIS-FFBC \cite{Tang}), it was assumed that the RIS always maintains the far-field condition (adequate distance) relative to both base stations and measurement points (more details in \cite{Tang}).

\subsection{Perfomance Metrics}
\label{subsection:performance_metrics}
To assess the improvement in radio signal path loss for the scenario utilizing RIS equipment compared to the standard scenario without any supporting devices like RIS matrices, the parameters defined by Eq.~(\ref{equation:pl_ris_dot})\textbf{--}(\ref{equation:pl_gain_avg}) were employed.

Eq.~(\ref{equation:pl_ris_dot}) and (\ref{equation:pl_no_ris_dot}) provide formulas to estimate the minimal path loss observed at a specific point on the map for the case with RIS enabled $\left(\dot{\text{PL}}_\text{RIS}\right)$ and the reference case with RIS deactivated $\left(\dot{\text{PL}}_\text{BS}\right)$. Additionally, Eq.~(\ref{equation:pl_avg_dot}) calculates the average path loss value, considering both the signal received \textit{directly} (via LOS or NLOS) from a base station and the signal reflected by the most efficient RIS matrix placement from the perspective of a given location.
\begin{align}
\label{equation:pl_ris_dot}
    &\dot{\text{PL}}_\text{RIS}\left(x,y\right)=\\ &\min_{i,j}\Bigl\{\text{PL}_{\text{LOS},i}\left(x,y\right);\text{}\text{PL}_{\text{NLOS},i}\left(x,y\right);\text{}\text{PL}_{\text{RIS},i,j}\left(x,y\right)\Bigr\}, \nonumber
\end{align}
\begin{equation}
\label{equation:pl_no_ris_dot}
    \dot{\text{PL}}_\text{BS}\left(x,y\right)=\min_{i}\Bigl\{\text{PL}_{\text{LOS},i}\left(x,y\right);\text{}\text{PL}_{\text{NLOS},i}\left(x,y\right)\Bigr\},
\end{equation}
\begin{equation}
\label{equation:pl_avg_dot}
    \dot{\text{PL}}_\text{AVG}\left(x,y\right)=\frac{\dot{\text{PL}}_\text{RIS}\left(x,y\right)+\dot{\text{PL}}_\text{BS}\left(x,y\right)}{2},
\end{equation}
where $x=0,\dots,X-1$ and $y=0,\dots,Y-1$ represent the coordinates of a specific path loss measurement point, and $X$ and $Y$ denote the total numbers of considered $x$ and $y$ coordinates, respectively. Similarly, $i=0,\dots,K_\text{BS}-1$ and $j=0,\dots,K_\text{RIS}-1$ are the identifiers for a base station and RIS device ($j$ dependent on placement), with $K_\text{BS}$ and $K_\text{RIS}$ representing the total number of deployed base stations and RIS locations. The terms $\text{PL}_{\text{LOS}}$, $\text{PL}_{\text{NLOS}}$, and $\text{PL}_{\text{RIS}}$ correspond to the radio signal path loss values received by mobile terminals \textit{directly} (via LOS), through obstacles causing signal distortion (via NLOS), and via reflections from RIS, respectively.

Before defining the final basic path loss parameter, let us clarify the additional factors included in Eq.~(\ref{equation:avg_trasmit_power}) and (\ref{equation:received_power}). The first equation represents the average transmit power $\left(\overline{P}_\text{TX}\right)$ per single network cell within the study area. This formula is derived as follows:
\begin{equation}
\label{equation:avg_trasmit_power}
    \overline{P}_{\text{TX}}=\frac{1}{K_\text{BS}}\sum_{i=0}^{K_\text{BS}-1}P_{\text{TX},i},
\end{equation}
where $P_{\text{TX},i}$ represents the transmit power for the $i$-th network cell $\left(i=0,\dots,K_\text{BS}-1\right)$. This average transmit power parameter is used in the subsequent equation to determine the power received at a specific point in the area $\left(P_\text{RX}\right)$. The formula for calculating this received power is as follows:
\begin{equation}
\label{equation:received_power}
    P_\text{RX}\left(x,y\right)=\overline{P}_{\text{TX}}-\dot{\text{PL}}\left(x,y\right).
\end{equation}

After introducing Eq.~(\ref{equation:avg_trasmit_power}) and (\ref{equation:received_power}), we can now define the final key PL parameter, which is crucial for evaluating the benefits of deploying the RIS device in the Radio Access Network area. The following equation defines the minimal radio signal path loss, assuming that at a particular measurement point, two signals \textbf{--} one transmitted via LOS or NLOS from a base station and the other reflected by the RIS \textbf{--} combine to reinforce each other $\left(\dot{\text{PL}}_\text{RIS,BS}\right)$:
\begin{align}
\label{equation:pl_ris_bs_dot}
    &\dot{\text{PL}}_\text{RIS,BS}\left(x,y\right)=\\ \nonumber &\overline{P}_{\text{TX}}-10\cdot\text{log}_{10}\Big(P_\text{RX,RIS}\left(x,y\right)+P_\text{RX,BS}\left(x,y\right)\Big),
\end{align}
where $P_\text{RX,RIS}$ and $P_\text{RX,BS}$ represent the signal powers received at a specific measurement point within the examined area, from a network cell via the RIS surface and through LOS or NLOS, respectively. Given that each network cell uses the same transmit power to propagate the radio signal, it can be assumed that $\overline{P}_\text{TX}=P_\text{TX}$.

The following equations outline the method for evaluating the parameters that represent the minimum $\left(\widetilde{\text{PL}}\right)$, maximum $\left(\widehat{\text{PL}}\right)$, and average $\Big(\overline{\text{PL}}\Big)$ path loss values. These values are calculated based on the estimated radio signal path loss distribution within the area under consideration.
\begin{equation}
\label{equation:pl_min}
    \widetilde{\text{PL}}=\min_{x,y}\left\{\dot{\text{PL}}\left(x,y\right)\right\},
\end{equation}
\begin{equation}
\label{equation:pl_max}
    \widehat{\text{PL}}=\max_{x,y}\left\{\dot{\text{PL}}\left(x,y\right)\right\},
\end{equation}
\begin{equation}
\label{equation:pl_avg}
    \overline{\text{PL}}=\frac{1}{XY}\sum^{Y-1}_{y=0}\sum^{X-1}_{x=0}\dot{\text{PL}}\left(x,y\right).
\end{equation}

Finally, the parameters described above are used to quantify the improvement in propagated radio signal strength within the examined area through the deployment of the RIS device in different locations. These gain metrics are expressed as a percentage in Eq.~(\ref{equation:pl_gain_min})\textbf{--}(\ref{equation:pl_gain_avg}), where the baseline is the path loss value obtained in the reference scenario (no RIS enabled \textbf{--} BS), and the key metrics are the values observed for specific scenarios: the minimal PL measured at a particular point with RIS enabled (RIS) and the average signal power received at a particular point by combining the \textit{direct} signal (via LOS or NLOS) and the optimal signal reflected by the RIS (RIS,BS).
\begin{equation}
\label{equation:pl_gain_min}
    G_\text{PL}^\text{min}=\scalebox{1.35}{\Bigg[}1-\frac{\left\{\widetilde{\text{PL}}_\text{RIS}\text{ }\lor\text{ }\widetilde{\text{PL}}_\text{RIS,BS}\right\}}{\widetilde{\text{PL}}_\text{BS}}\scalebox{1.35}{\Bigg]} \cdot 100\text{ }\%,
\end{equation}
\begin{equation}
\label{equation:pl_gain_max}
    G_\text{PL}^\text{max}=\scalebox{1.35}{\Bigg[}1-\frac{\left\{\widehat{\text{PL}}_\text{RIS}\text{ }\lor\text{ }\widehat{\text{PL}}_\text{RIS,BS}\right\}}{\widehat{\text{PL}}_\text{BS}}\scalebox{1.35}{\Bigg]} \cdot 100\text{ }\%,
\end{equation}
\begin{equation}
\label{equation:pl_gain_avg}
    G_\text{PL}^\text{avg}=\scalebox{1.35}{\Bigg[}1-\frac{\left\{\overline{\text{PL}}_\text{RIS}\text{ }\lor\text{ }\overline{\text{PL}}_\text{RIS,BS}\right\}}{\overline{\text{PL}}_\text{BS}}\scalebox{1.35}{\Bigg]} \cdot 100\text{ }\%.
\end{equation}

\subsection{Results Overview}
\label{subsection:results_overview}
Tab.~\ref{table:results_pl_values} presents the minimal, maximal, and average path loss values for various mobile system scenarios. These scenarios include measurements of propagation attenuation for radio signals transmitted solely by base stations via LOS or NLOS, depending on the BS-MP visibility (BS), signals transmitted \textit{directly} by base stations (via LOS or NLOS) and through the RIS (RIS), and scenarios where those radio signals are reinforced (RIS,BS). Additionally, the table includes the average path loss values when the signal can be received from either RIS or BS (AVG).
\begin{figure*}[htbp]
\centering
\includegraphics[width=0.99\textwidth]{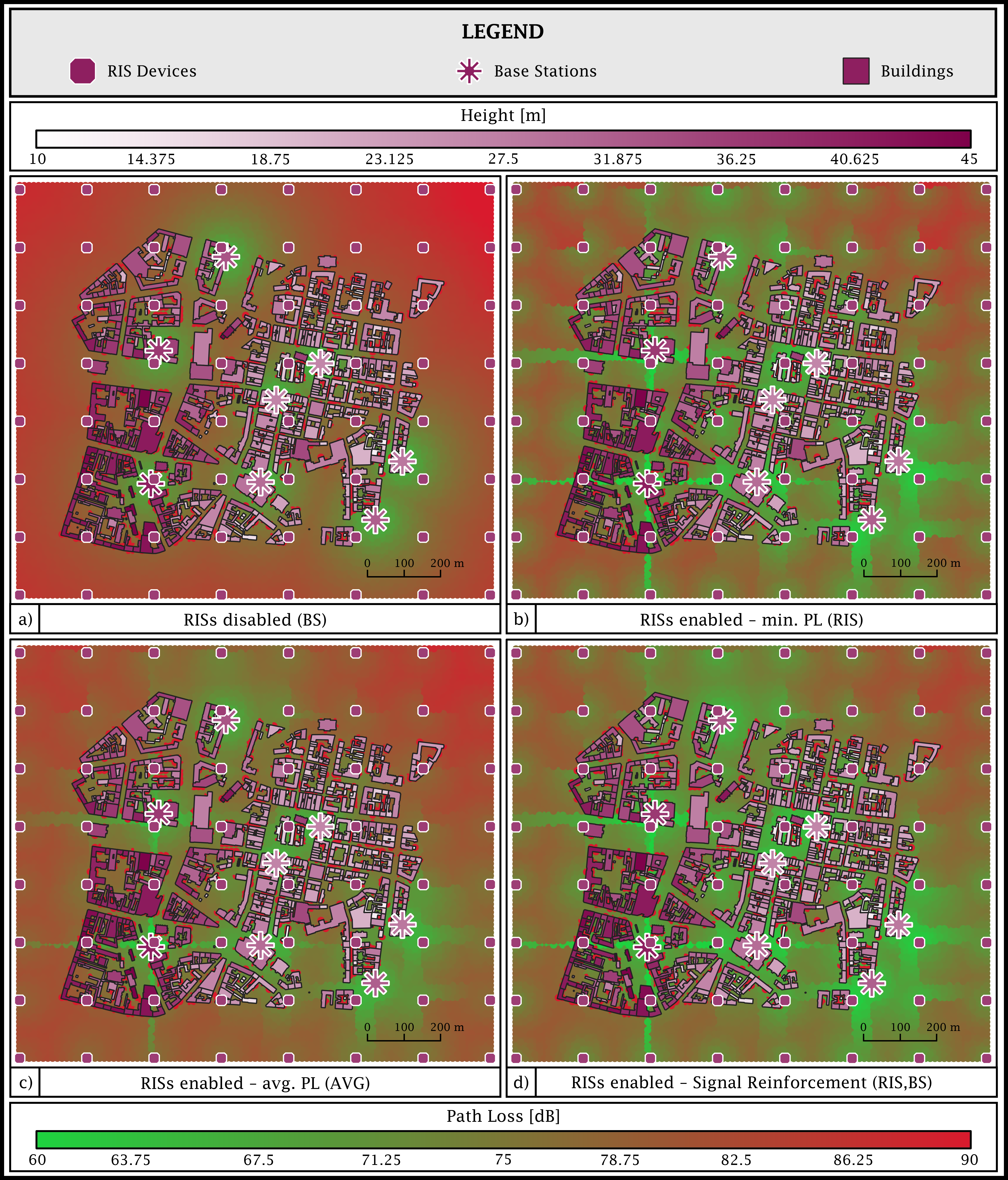}
\caption{Path Loss distribution for BS (a), RIS (b), AVG (c), and RIS,BS (d) scenarios}
\label{figure:pl_distribution}
\end{figure*}

The parameters presented in Tab.~\ref{table:results_pl_values} were collected based on measurements taken across the entire examined area. Fig.~\ref{figure:pl_distribution}.a-d illustrate the path loss distribution for the system, highlighting results for all four approaches (a \textbf{--} BS, b \textbf{--} RIS, c \textbf{--} AVG, d \textbf{--} RIS,BS). Each figure features two color scales placed above and below the included maps. The scale at the top indicates the height of objects within the area ($\leq10$ m in white, $\geq45$ m in purple), providing context for how buildings and other obstacles impact the path loss measurements. The bottom scale represents the path loss value at specific measurement points ($\leq60$ dB in green, $\geq90$ dB in red). These color scales enable a visual assessment of potential reductions in radio signal attenuation within the area.

Additionally, the parameters in Tab.~\ref{table:results_pl_values} were used to evaluate the gains (Tab.~\ref{table:results_gain}) associated with reducing path loss for each scenario involving RIS deployment (RIS, RIS,BS, and AVG) compared to the scenario with RIS disabled (BS).

Now, let's review the results summarized in Tab.~\ref{table:results_pl_values} and \ref{table:results_gain}, as well as Fig.~\ref{figure:pl_distribution}. Compared to the reference scenario (BS, shown in Fig.~\ref{figure:pl_distribution}.a) where the RIS matrix is disabled, the highest gain is observed in the signal reinforcement approach (RIS,BS, shown in Fig.~\ref{figure:pl_distribution}.d). For the reference scenario, the observed path loss values were as follows: a minimum of $57.46$ dB, a maximum of $91.78$ dB, and an average of $77.13$ dB. In contrast, for the signal reinforcement approach (RIS,BS), the values were $42.67$ dB, $84.85$ dB, and $70.31$ dB, respectively. This corresponds to a gain of $25.73\text{ }\%$ for minimum PL, $7.55\text{ }\%$ for maximum PL, and $8.84\text{ }\%$ for average PL compared to the initial reference scenario (BS).

The next best performance is seen in the minimum PL selection approach (RIS, shown in Fig.~\ref{figure:pl_distribution}.b), where the RIS device is activated. The recorded values were $42.71$ dB (min.), $87.62$ dB (max.), and $71.68$ dB (avg.). These results correspond to gains of $25.66\text{ }\%$ for minimum PL, $4.54\text{ }\%$ for maximum PL, and $7.06\text{ }\%$ for average PL compared to the scenario without RIS device.

Finally, the technique that averages radio signal attenuation received from both RIS and BSs (AVG, shown in Fig.~\ref{figure:pl_distribution}.c) shows the least improvement, with gains of $8.42\text{ }\% $for minimum PL, $3.27\text{ }\%$ for maximum PL, and $3.53\text{ }\%$ for average PL (min.: $52.62$ dB, max.: $88.79$ dB, avg.: $74.4$ dB).
\begin{table}[ht]
\centering
\caption{Impact of RIS enabling on PL distribution}
\label{table:results_pl_values}
\resizebox{0.49\textwidth}{!}{
\begin{tabular}{|c|c|c|c|c|}
\arrayrulecolor[HTML]{002060}
\hline
\multirow{3}{*}{\cellcolor[HTML]{002060}{\color[HTML]{FFFFFF} }} & \multicolumn{4}{c|}{\multirow{1}{*}{\cellcolor[HTML]{002060}{\color[HTML]{FFFFFF} Value}}} \\ \cline{2-4}
\multirow{3}{*}{\cellcolor[HTML]{002060}{\color[HTML]{FFFFFF} }} & \cellcolor[HTML]{FFFFFF}{\color[HTML]{000000} \textit{BSs}} & \cellcolor[HTML]{FFFFFF}{\color[HTML]{000000} \textit{BSs $\lor$ RIS}} & \cellcolor[HTML]{FFFFFF}{\color[HTML]{000000} \textit{BSs + RIS}} & \cellcolor[HTML]{FFFFFF}{\color[HTML]{000000} \textit{AVERAGE}} \\ 
\multirow{-3}{*}{\cellcolor[HTML]{002060}{\color[HTML]{FFFFFF} Parameter}} & 
\cellcolor[HTML]{FFFFFF}{\color[HTML]{000000} (BS)} & \cellcolor[HTML]{FFFFFF}{\color[HTML]{000000} (RIS)} & \cellcolor[HTML]{FFFFFF}{\color[HTML]{000000} (RIS,BS)} & \cellcolor[HTML]{FFFFFF}{\color[HTML]{000000} (AVG)} \\ \hline
\multirow{-1.5}{*}{\cellcolor[HTML]{E7E7E7}{\color[HTML]{000000} $\widetilde{\text{PL}}$}}\rule{0pt}{3ex} & \multirow{-1.5}{*}{\cellcolor[HTML]{E7E7E7}{\color[HTML]{000000} $57.46\text{ dB}$}} & \multirow{-1.5}{*}{\cellcolor[HTML]{E7E7E7}{\color[HTML]{000000} $42.71\text{ dB}$}} & \multirow{-1.5}{*}{\cellcolor[HTML]{E7E7E7}{\color[HTML]{000000} $42.67\text{ dB}$}} & \multirow{-1.5}{*}{\cellcolor[HTML]{E7E7E7}{\color[HTML]{000000} $52.62\text{ dB}$}} \\ \hline
\multirow{-1.5}{*}{\cellcolor[HTML]{FFFFFF}{\color[HTML]{000000} $\widehat{\text{PL}}$}}\rule{0pt}{3ex} & \multirow{-1.5}{*}{\cellcolor[HTML]{FFFFFF}{\color[HTML]{000000} $91.78\text{ dB}$}} & \multirow{-1.5}{*}{\cellcolor[HTML]{FFFFFF}{\color[HTML]{000000} $87.62\text{ dB}$}} & \multirow{-1.5}{*}{\cellcolor[HTML]{FFFFFF}{\color[HTML]{000000} $84.85\text{ dB}$}} & \multirow{-1.5}{*}{\cellcolor[HTML]{FFFFFF}{\color[HTML]{000000} $88.79\text{ dB}$}} \\ \hline
\multirow{-1.5}{*}{\cellcolor[HTML]{E7E7E7}{\color[HTML]{000000} $\overline{\text{PL}}$}}\rule{0pt}{3ex} & \multirow{-1.5}{*}{\cellcolor[HTML]{E7E7E7}{\color[HTML]{000000} $77.13\text{ dB}$}} & \multirow{-1.5}{*}{\cellcolor[HTML]{E7E7E7}{\color[HTML]{000000} $71.68\text{ dB}$}} & \multirow{-1.5}{*}{\cellcolor[HTML]{E7E7E7}{\color[HTML]{000000} $70.31\text{ dB}$}} & \multirow{-1.5}{*}{\cellcolor[HTML]{E7E7E7}{\color[HTML]{000000} $74.4\text{ dB}$}} \\ \hline
\end{tabular}}
\end{table}
\begin{table}[ht]
\centering
\caption{PL distribution gain caused by the use of RIS}
\label{table:results_gain}
\resizebox{0.39\textwidth}{!}{
\begin{tabular}{|c|c|c|c|}
\arrayrulecolor[HTML]{002060}
\hline
\multirow{3}{*}{\cellcolor[HTML]{002060}{\color[HTML]{FFFFFF} }} & \multicolumn{3}{c|}{\multirow{1}{*}{\cellcolor[HTML]{002060}{\color[HTML]{FFFFFF} Value}}} \\ \cline{2-3}
\multirow{3}{*}{\cellcolor[HTML]{002060}{\color[HTML]{FFFFFF} }} & \cellcolor[HTML]{FFFFFF}{\color[HTML]{000000} \textit{BSs $\lor$ RIS}} & \cellcolor[HTML]{FFFFFF}{\color[HTML]{000000} \textit{BSs + RIS}} & \cellcolor[HTML]{FFFFFF}{\color[HTML]{000000} \textit{AVERAGE}} \\ 
\multirow{-3}{*}{\cellcolor[HTML]{002060}{\color[HTML]{FFFFFF} Parameter}} & \cellcolor[HTML]{FFFFFF}{\color[HTML]{000000} (RIS)} & \cellcolor[HTML]{FFFFFF}{\color[HTML]{000000} (RIS,BS)} & \cellcolor[HTML]{FFFFFF}{\color[HTML]{000000} (AVG)} \\ \hline
\multirow{-1.5}{*}{\cellcolor[HTML]{E7E7E7}{\color[HTML]{000000} $G_\text{PL}^\text{min}$}}\rule{0pt}{3ex} & \multirow{-1.5}{*}{\cellcolor[HTML]{E7E7E7}{\color[HTML]{000000} $25.66\text{ }\%$}} & \multirow{-1.5}{*}{\cellcolor[HTML]{E7E7E7}{\color[HTML]{000000} $25.73\text{ }\%$}} & \multirow{-1.5}{*}{\cellcolor[HTML]{E7E7E7}{\color[HTML]{000000} $8.42\text{ }\%$}} \\ \hline
\multirow{-1.5}{*}{\cellcolor[HTML]{FFFFFF}{\color[HTML]{000000} $G_\text{PL}^\text{max}$}}\rule{0pt}{3ex} & \multirow{-1.5}{*}{\cellcolor[HTML]{FFFFFF}{\color[HTML]{000000} $4.54\text{ }\%$}} & \multirow{-1.5}{*}{\cellcolor[HTML]{FFFFFF}{\color[HTML]{000000} $7.55\text{ }\%$}} & \multirow{-1.5}{*}{\cellcolor[HTML]{FFFFFF}{\color[HTML]{000000} $3.27\text{ }\%$}}  \\ \hline
\multirow{-1.5}{*}{\cellcolor[HTML]{E7E7E7}{\color[HTML]{000000} $G_\text{PL}^\text{avg}$}}\rule{0pt}{3ex} & \multirow{-1.5}{*}{\cellcolor[HTML]{E7E7E7}{\color[HTML]{000000} $7.06\text{ }\%$}} & \multirow{-1.5}{*}{\cellcolor[HTML]{E7E7E7}{\color[HTML]{000000} $8.84\text{ }\%$}} & \multirow{-1.5}{*}{\cellcolor[HTML]{E7E7E7}{\color[HTML]{000000} $3.53\text{ }\%$}} \\ \hline
\end{tabular}
}
\end{table}

\begin{figure*}[htp]
\centering
\includegraphics[width=\textwidth]{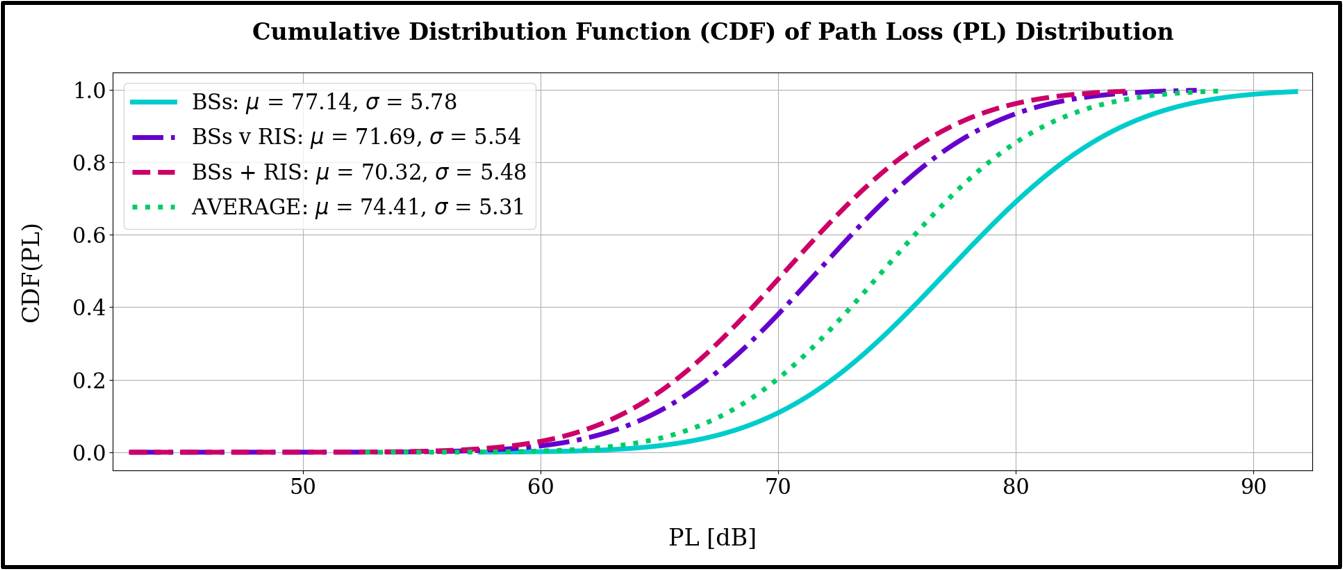}
\caption{CDF plots of the Path Loss distribution for all scenarios (BS \textbf{--} blue line, RIS \textbf{--} purple line, RIS,BS \textbf{--} red line, AVG \textbf{--} green line)}
\label{figure:cdf_pl}
\end{figure*}

The conclusions drawn from the study are further illustrated and reinforced by the plots presented in Fig.~\ref{figure:cdf_pl}, which depict the Cumulative Distribution Function for each path loss measurement approach. These CDF plots provide a comprehensive view of the distribution and variability of path loss across the different scenarios, allowing for a clearer comparison of the effectiveness of RIS deployment in improving signal propagation. By analyzing the CDF curves, we can observe the proportion of measurement points that experience specific levels of path loss, thereby offering deeper insights into the overall performance and benefits of each approach. This visualization underscores the practical implications of the study's findings, highlighting the scenarios where RIS technology can most significantly enhance network performance.

\section{Conclusions}
\label{section:conclusions}
The study presents an analysis of radio signal path loss values under different mobile system scenarios involving base stations and Reconfigurable Intelligent Surfaces. The results indicate that the signal reinforcement approach (RIS,BS) provides the highest improvement in reducing path loss, with significant gains in minimal, maximal, and average PL values compared to the baseline scenario where RIS is disabled (BS). The RIS-considering (RIS) scenario also shows considerable improvement, though slightly less effective than the signals amplifying (RIS,BS) approach. The PL averaging technique (AVG) offers the least improvement in signal attenuation reduction. These findings are further illustrated by the CDF plots, confirming the effectiveness of RIS engagement in mobile systems placed in urban environments.

Although the research demonstrates significant improvements in signal attenuation reduction, certain simplifications were made to showcase the maximal path loss (PL) gains for the system scenario under consideration. Specifically, fixed values for azimuth and elevation angles were assumed, which may not always be achievable during continuous mobile service delivery to moving users within the studied area. For future work, it is essential to consider these dynamic variations to more accurately assess the impact of RISs in realistic, real-world use cases.

\section*{Acknowledgements}
\label{section:ackonwledgements}
The authors express their gratitude to Prof. Margot Deruyck from Ghent University (IMEC) in Belgium for her invaluable support of this work through the provision of the GRAND software, which significantly enhanced the research outcomes.

Adrian Kliks is also affiliated with the Luleå University of Technology in Sweden, where he contributes to research and academic initiatives.


\begin{thebibliography}{00}
\bibitem{Huang} C. Huang, A. Zappone, G.C. Alexandropoulos, M.~Debbah, and C. Yuen, "Reconfigurable Intelligent Surfaces for Energy Efficiency in Wireless Communication," {\it IEEE Transactions on Wireless Communications}, vol. $18$, no. $8$, pp. $4157$\textbf{–-}$4170$, $2019$. DOI: $10$.$1109$/TWC.$2019$.$2922609$.

\bibitem{DiRenzo} M. Di Renzo et al., "Smart Radio Environments Empowered by Reconfigurable Intelligent Surfaces: How It Works, State of Research, and The Road Ahead," {\it IEEE Journal on Selected Areas in Communications}, vol. $38$, no. $11$, pp. $2450$\textbf{--}$2525$, $2020$. DOI: $10$.$1109$/JSAC.$2020$.$3007211$.

\bibitem{Tang} W. Tang et al., "Wireless Communications With Reconfigurable Intelligent Surface: Path Loss Modeling and Experimental Measurement," {\it IEEE Transactions on Wireless Communications}, vol. $20$, no. $1$, pp. $421$\textbf{--}$439$, $2021$. DOI: $10$.$1109$/TWC.$2020$.$3024887$.

\bibitem{Liu} Y. Liu et al., "Reconfigurable Intelligent Surfaces: Principles and Opportunities," {\it IEEE Communications Surveys \& Tutorials}, vol. $23$, no. $3$, pp. $1546$\textbf{--}$1577$, $2021$. DOI: $10$.$1109$/COMST.$2021$.$3077737$.

\bibitem{SamorzewskiJTIT2023} A. Samorzewski, "Energy Consumption in Wireless Systems Equipped with RES, UAVs, and IRSs," {\it Journal of Telecommunications and Information Technology}, no. $2$, pp. $35\textbf{}${--}$40$, $2023$. DOI: $10$.$26636$/jtit.$2023$.$170923$.

\bibitem{SamorzewskiKRiT2023} A. Samorzewski and A. Kliks "5G cellular systems supported by UAVs, RESs, and RISs," in {\it Radiocommunication and Teleinformatics Conference 2023 (pol. Konferencja Radiokomunikacji i Teleinformatyki 2023 -- KRiT 2023)}, Cracow, Poland, $2023$, pp.~$97\textbf{}${--}$100$. DOI: $10$.$15199$/$59$.$2023$.$4$.$18$.

\bibitem{SamorzewskiSoftCOM2023} A. Samorzewski and A. Kliks, "5G Networks Supported by UAVs, RESs, and RISs," in {\it 2023 International Conference on Software, Telecommunications and Computer Networks (SoftCOM)}, Split, Croatia, $2023$, pp.~$1$\textbf{--}$6$. DOI: $10$.$23919$/SoftCOM$58365$.$2023$.$10271683$.

\bibitem{SamorzewskiGLOBECOM2023} A. Samorzewski, M. Deruyck, and A. Kliks, "Energy Consumption in RES-Aware 5G Networks," in~{\it GLOBECOM 2023 -- 2023 IEEE Global Communications Conference}, Kuala Lumpur, Malaysia, $2023$, pp. $1024\textbf{}${--}$1029$. DOI: $10$.$1109$/GLOBECOM$54140$.$2023$.$10437451$.

\bibitem{AreaData} {\it Poznan \textbf{--} Model 3D}. SIP (ang. Spatial Information System). [Online]. Available: http://sip.poznan.pl/model3d/\#/legend.

\bibitem{NetworkData} {\it Database and map of BTS station locations / UKE permits}. BTSearch. [Online]. Available: http://beta.btsearch.pl.

\bibitem{Castellanos} G. Castellanos, S. De Gheselle, L. Martens, N. Kuster, W. Joseph, M. Deruyck, and S. Kuehn, "Multi-objective optimization of human exposure for various $5$G network topologies in Switzerland," {\it Computer Networks}, vol. $2016$, $2022$. DOI: $10$.$1016$/j.comnet.$2022$.$109255$.

\bibitem{3GPP} $3$GPP, "Technical Specification Group Radio Access Network; Study on channel model for frequencies from $0.5$ to $100$ GHz (Release $18$)," TR $38.901$ v$18.0.0$, $2024$.

\bibitem{Bjornson} E. Bj{\"o}rnson, J. Hoydis, and L. Sanguinetti, "Massive MIMO Networks: Spectral, Energy, and Hardware Efficiency," {\it Foundations and Trends® in Signal Processing}, vol. $11$, no. $3$\textbf{--}$4$, pp. $154$\textbf{--}$655$, $2017$. DOI: $10$.$1561$/$2000000093$.
\end{thebibliography}
\end{document}